\documentstyle[prd,aps]{revtex}

\newcommand{\bea}{\begin{eqnarray}}
\newcommand{\eea}{\end{eqnarray}}
\newcommand{\be}{\begin{equation}}
\newcommand{\ee}{\end{equation}}

\begin{document}
%%%%%%%%%%%%%%%%%%%%%%%%%%%%%%%%%%%%%%%%%%%%%%%%%%%%%%%%%%%%%%%
\draft
%  For 2 column format.
\twocolumn[\hsize\textwidth\columnwidth\hsize\csname
@twocolumnfalse\endcsname

%%%%%%%%%%%%%%%%%%%%%%%%%%%%%%%%%%%%%%%%%%%%%%%%%%%%%%%%%%%%%%%
\title{Novel Symmetries in Axion-Dilaton String Cosmology }
\author{Jnanadeva Maharana${}^{(a,b)}$ \\
        ${}^{(a)}$Theory Division,  
                   KEK, Tsukuba, Ibaraki 305-0801, Japan  \\
        ${}^{(b)}$ Institute of Physics, Bhubaneswar-751 005, India}
%\date{\today}
\maketitle

%%%%%%%%%%%%%%%%%%%%%%%%%%%%%%%%%%%%%%%%%%%%%%%%%%%%%%%%%%%%%%%
\begin{abstract}

The symmetry structure of axion-dilaton quantum string cosmology is
investigated. The invariance of the string effective action under
S-duality group, $SU(1,1)$, facilitates solution of Wheeler-De Witt
equation from group theoretic considerations; revealing existence 
of a new class of wave functions.  We discover the an underlying 
 ${W}$-infinity algebra in this formulation.  

\end{abstract}

%%%%%%%%%%%%%%%%%%%%%%%%%%%%%%%%%%%%%%%%%%%%%%%%%%%%%%%%%%%%%%%
%  For 2 column format.
\vskip2pc]

%%%%%%%%%%%%%%%%%%%%%%%%%%%%%%%%%%%%%%%%%%%%%%%%%%%%%%%%%%%%%%%
\noindent It is recognized that symmetries
 play a very important role in understanding
diverse properties of string theories. Dualities provide  foundations to
study string dynamics and to  construct  a unique and unified theory
 \cite{js1}. It is 
well known that string effective actions in lower spacetime dimensions tend to
exhibit higher symmetries. For example, the 4-dimensional heterotic string
effective action derived from $D=10$ action by toroidal compactification 
on $T^6$,
is endowed with $O(6,22)$ T-duality symmetry. The 3-form field strength
may be dualized to introduce the pseudoscalar axion, $\chi$. The dilaton,
$\phi$, and $\chi$ parametrize the coset ${SL(2,R)}\over {U(1)}$ $\sim$
${SU(1,1)}\over {U(1)}$. Thus the duality group \cite{sen1} of $D=4$ theory
is $O(6,22) \times SU(1,1)$. If we envisage the theory in $D=3$ by compactifying
the $D=10$ theory on $T^7$, then an  enlarged group,  $O(8,32)$, 
emerges as the duality group. The two dimensional effective action derived
through dimensional reduction on $T^8$ incorporates affine algebra\cite{jm}.\\ 
The string cosmological scenario, where backgrounds depend on cosmic time,
is quite interesting   
in $D=d+1$ dimensions. 
The dilaton-graviton action is invariant under
 the scale factor duality \cite{gv1} which
is a subgroup of $O(d,d)$ group,
 d being number of spatial dimensions. The O(d,d) symmetry appears \cite{kmgv}  
if the time dependent B-field is incorporated in 
the action. There has been considerable amount of 
activities to study cosmologies
in the frame work string theory and M-theory in recent years \cite{cop1,gv2}. \\
The pair $(\phi , \chi)$ play a special role among a large number
of massless scalars, present in 4-dimensional string theories. Whereas,
$e^{<\phi >}$  determines $G_N , ~\alpha _{YM}$ and other  coupling constants
of the theory; $\chi$ is identified with
the elusive axion. Although, it was introduced in the context of QCD,
axion is recognized to play a  significant role in cosmology.\\
In this letter we study isotropic, homogeneous 
dilaton-axion string cosmology and
explore its symmetry content. We present a  general class of solutions to the
Wheeler-De Witt (WDW) equation. The
S-duality invariance of the action is exploited to solve WDW equation
from the group theoretic considerations. The wave function is factorized 
as a product of a function of 
the scale factor and an eigenfunction of the Casimir of
$SU(1,1)$, the S-duality group.  These wave functions  
are  new and were missed earlier
works \cite{mmp}.  
It is argued that one can  construct a set of operators from the  
 generators of $ SU(1,1)$  
and show that these 
operators close into an infinite dimensional algebra. 
We show that the
$w_{\infty}$ algebra proposed by Bakas 
\cite{bakas} emerges in a specific example.   
In recent years $w_{\infty}$ and $W_{\infty}$, which might
be considered as deformation of the former,  have played useful roles in 
a variety of problems 
in physics. They first appeared  as $N\rightarrow \infty$ limit
of $W_N$ algebras \cite{wn}. These algebras were studied in the context of $c=1$
theories \cite{c1}, in large N limit of $SU(N)$gauge theories \cite{g1}
 and in quantum Hall effect \cite{qh}. It
is quite surprizing that such algebras now resurfaces in string cosmology.  
In the past, symmetry content of string
cosmologies have been explored from diverse perspectives \cite{sqs} and our
efforts are in yet another novel direction.\\
Let us consider an action in the Einstein frame 
\bea
\label{action1}
S=\int d^4x{\sqrt{-g}}\left( R-{1\over 2}\partial_{\mu}\phi \partial ^{\mu}\phi
  -{1\over 2}
e^{2\phi}\partial _{\mu}\chi \partial ^{\mu}\chi  \right)
\eea
where $g_{\mu\nu}$ is the metric,  $g$ its determinant, $\mu ,\nu =0,1,2,
3$ and  R is the 
scalar curvature. 
One can derive such an action from four
dimensional heterotic string theory, after dualising 
$H^{\mu\nu\rho}=e^{2\phi}{\epsilon
^{\mu\nu\rho\lambda}\partial _{\lambda}\chi}$ and 
 setting rest of the backgrounds to
zero or by compactifying \cite{jm2}
 $D=10$ type IIB theory on $T^6$ and truncating
action to the above form. 
One may reexpress (\ref{action1})
in a manifestly S-duality, $SU(1,1)$,  invariant form
 \bea
\label{su11}
S=\int d^4x{\sqrt{-g}}\left(R +{1\over 4}
{\rm Tr}[\partial _{\mu}{\bf V}^{-1}(x)
\partial ^{\mu} {\bf V}(x)]\right) 
\eea
where
\bea
\label{mxv}
{\bf V} ={1\over 2}\pmatrix {A+B & 2B\chi +i(A-B) \cr 2B\chi -i(A-B) & A+B \cr}
\eea
The elements of the $2\times 2$
matrix $\bf V$ are defines as: $A=e^{-\phi}+\chi ^2e^{\phi}
$ and  $B=e^{\phi}$. Note that ${\bf V} \in SU(1,1)$ and satisfies 
${\bf V}^{-1}=\sigma _3{\bf V}^{\dagger}\sigma _3$; 
$\sigma _i ,i=1,2,3$
denote the Pauli matrices. The action (\ref{su11}) is invariant under
\bea
{\bf V}\rightarrow \Omega ^{\dagger}{\bf V}\Omega, ~~~g_{\mu\nu}\rightarrow
g_{\mu\nu} ~{\rm and}~~\Omega ^{\dagger}\sigma _3 \Omega =\sigma _3,
\eea
where $\Omega \in SU(1,1)$
 and $\sigma _3$ is the SU(1,1) metric which remains invariant under the 
transformation. We mention {\it en passant } that  the action is usually 
expressed in $SL(2,R)$ invariant form in terms of the 
${\cal M}$-matrix \cite{js2}.
The two matrices are related, $i.e.$  ${\cal M}=e^{-i{{\sigma _1\pi}\over 4}}
{\bf V}
e^{i{{\sigma _1\pi}\over 4}}$.
Note that any $SU(1,1)$ matrix can be expressed as
\bea
\label{linear}
{\bf V}={\bf 1}v_0+\sum _{i=1}^3 \sigma _iv_i
\eea
${\bf 1}$ being the unit matrix and the spacetime dependent 
coefficients satisfy:
$v_0^2-v_1^2-v_2^2+v_3^2=1 $. 
For ${\bf V}$ given by (\ref{mxv}), $v_3=0$; therefore, the constraint
reads
\bea
\label{cv1}
v_0^2-v_1^2-v_2^2=1
\eea
with $v_0={1\over 2}(A+B)$, $v_1=B\chi $ and $v_2={1\over 2}(B-A)$ 
which satisfy the desired constraint (\ref{cv1}). In fact  
(\ref{cv1}) defines a space of constant Gaussian curvature 
 in the {\it moduli}  space. \\
In the cosmological context, the  FRW spacetime 
metric  is
\bea
\label{frw}
ds^2=-dt^2+a(t)^2\left({{dr^2}\over {1-kr^2}}+r^2d\Omega ^2 \right)
\eea
$a(t)$ being the scale factor; $k=+1,0,-1$ correspond to closed, flat
and open Universes respectively. Furthermore, the $\bf V$-matrix or
 alternatively, $v_0, v_1,v_2$ depend on cosmic time, $t$. From
now on, we consider,  $k=1$, closed Universe. The  scalar curvature
derived from (\ref{frw}) is $R=6(-a{\dot a}^2 +a)$, up to
a total derivative term. Here and everywhere overdot stands for
 ${d\over {dt}}$. 
 It is convenient to scale the metric to get rid of the
 factor 6 in the expression for the curvature scalar and the redefine  
 the  matter fields 
suitably so that action takes  the following form:
\bea
\label{newact}
S={1\over 2}\int dt\left(-a{\dot a}^2+a +a^3{\dot v}_i{\dot v}_j\eta^{ij}
 \right)
\eea
 The metric $\eta ^{ij} ={\rm diag}(1,-1,-1)
$ and the constraint is: $\eta ^{ij}v_iv_j=1$.
 The three conserved quantities; 
$ X_1=(v_1{\dot v}_2 - v_2{\dot v}_1), ~X_2=(v_0{\dot v}_2-v_2{\dot v}_0)$ and 
$X_3=(v_1{\dot v}_0-v_0{\dot v}_1)$, constructed from (\ref{newact})  
satisfy $SU(1,1)$ algebra. Note  that $X_1$ is  compact and generates
rotation around $v_0$ axis and other two correspond to  boosts. We recall that
the noncompact group $SU(1,1)$ is endowed with three generators: 
$J_i , i=1,2,3$  and they satisfy the following algebra:
\bea 
\label{algebra}
[J_1,J_2]=-iJ_3, ~~~[J_2,J_3]=iJ_1, ~~~[J_3,J_1]=iJ_2 
\eea
The raising/lowering operators: $J_{\pm}=J_1\pm iJ_2$
obey the commutation relations: $[J_3,J_{\pm}]=\pm J_{\pm}$ 
and $[J_{+}, J_{-}]=-2J_3$.
The Casimir operator is  
$C=-J_1^2-J_2^2+J_3^2$.  \\
The Hamiltonian constraint  reads
\bea
\label{hamiltonian}
H={1\over 2}\left(-{1\over a}P_a^2-a+{{1\over {a^3}}}\eta _{ij}P^i_vP^j_v 
\right) =0.
\eea
where $H$ is the Hamiltonian derived from (\ref{newact}) 
and  $P_a$ and $P^i_v$ are canonical momenta associated with $a$  
and $v_i$ respectively. In quantum cosmology, the Hamiltonian constraint 
(\ref{hamiltonian}) translates into WDW equation. The Hamiltonian operator
$\hat H$ is defined   
by identifying 
 ${\hat P}_a=-i{{\partial}
\over {\partial a}}$ and ${\hat P}^i_{v}=-i{{\partial}\over {\partial v_i}}$.
 The standard
procedure is to  
derive the wave function for a given $\hat H$, satisfying (\ref{hamiltonian}).   
 Notice, however that the scale factor
remains unchanged under $SU(1,1)$ transformation and a little calculation
shows that the term $\eta _{ij}P^i_vP^j_v$ in (\ref{hamiltonian}) 
is the Casimir operator of 
the S-duality group.  The problem is analogous to that of motion of a particle
in spherically symmetric potential, where the $L^2$ term contributes the
angular momentum barrier  leaving one to solve for the radial equation. The
quantum Hamiltonian needs to be defined with an operator ordering prescription
as is obvious from inspection of (\ref{hamiltonian}). 
We adopt the one \cite{swh} which
respects coordinate invariance in the minisuperspace and the WDW equation takes 
form 
\bea
\label{wdw}
\left({{{\partial}^2}\over {\partial a^2}}+{{\partial}\over{\partial a}}-a^2
+{1\over a^2}C\right){\Psi}=0 . 
\eea 
 We may factorize the wave function $\Psi ={\cal U}(a) Y$ where Y
is the eigenfunction of the Casimir operator satisfying:
 $CY=j(j+1)Y$. Next, we seek simultaneous eigenfunction of $C$ and the
compact generator $J_3$ and 
 exploit the well known group theoretic results \cite{group} of $SU(1,1)$. 
The wave functions satisfy the requirements that
\bea
C|j,m>=j(j+1)|j,m>,~ {\rm and } ~J_3|j,m>=m|j,m> 
\eea
We need the  unitary infinite
 dimensional representations 
 of $SU(1,1)$, to solve for the `angular' part of 
WDW equation, which are classified as follows. \\
(i) The discrete series $D^{\pm}_j$: $D^{+}_j$ is defined such that $j=-1/2, -1,
-3/2, -2,$... and for a given $j$, $m$ is unbounded
from above i.e. $m=-j, -j+1, -j+2,...$. The other discrete series, $D^{-}_j$
for which j is negative (integer or half integer), is the one where $m$
is unbounded from below i.e. $m=j, j-1, j-2,..$. There is a symmetry between
these sets of wave functions under $m\rightarrow -m$ which follows from the
properties of the $d^j_{mm'}$ functions of $SU(1,1)$. Note that $j$ is negative
in our convention.\\
(ii) The continuous series $C^0_k$ and $C^{1/2}_k$ correspond to 
\bea
j=-{1\over 2}+ik, ~~k>0,~~{\rm and ~ real}
\eea
with $m=0,\pm 1, \pm 2,..$  and $m=\pm 1/2, \pm 3/2,..$ for $C^0_k$ and 
$C^{1/2}_k$ respectively; also note that $j^{*} =-j-1$. Since $k$ takes
continuous values, some times these states are called `scattering states'
in the literature.\\
(iii) The supplementary series is defined for  $-1/2<j<0$ and $m=0,
\pm 1,\pm 2...$. It is well known that the set $\{D^{\pm}_j,C^0_k,C^{1/2}_k \}$ 
provide a complete set of basis functions. Therefore, any
function corresponding to ${SU(1,1)}\over {U(1)}$ may  be expanded in this
basis. One may omit the supplementary series while looking for the wave
functions from purely group theoretic considerations. However, we shall utilize 
this series in an  example.\\
Let us focus attention on solution of the WDW equation (\ref{wdw}). 
We define a `polar' coordinate system: $v_0={\rm cosh}\alpha ,v_1={\rm sinh}
\alpha {\rm cos}\beta$ and $v_2={\rm sinh}\alpha {\rm sin}\beta$; $\alpha$ real
and $0\le \beta \le 4\pi$. 
The Casimir of the $SU(1,1)$ is given by the Lapace-Beltrami operator \cite{bi}
\bea
\label{lb}
-{1\over {{\rm sinh}\alpha}}{{\partial}\over {\partial \alpha}}
{{\rm sinh \alpha}}
{{\partial}\over {\partial \alpha}}-{1\over {{\rm sinh}^2\alpha}}{{{\partial}^2}
\over {\partial {\cal \beta}^2}}
\eea
and the wave function $Y$, satisfying:  $CY=j(j+1)Y$ is given by
$Y^m_j({\rm cosh}\alpha ,\beta)=e^{im\beta}P^m_j({\rm cosh}\alpha)$. 
It is eigenfunction of both $C$ and compact generator 
 $J_3$. The  `magnetic' quantum number, $m$, is quantized and   
$P^m_j({\rm cosh}\alpha)$ are associated Lengedre polynomials. \\
Now we proceed to solve the equation for ${\cal U}(a)$
\bea
\label{scalef}
\left({{d^2}\over {d a^2}}+{{d}\over {da}} -a^2 +{{j(j+1)}\over {a^2}}\right)
{\cal U}(a)=0
\eea
The solutions are Bessel functions \cite{bessel}:
${\cal U}(a)_{\nu}=J_{\pm i{\nu}/2}({i\over 2}a^2) $
where ${\nu}^2=j(j+1)$ is introduced for notational conveniences.Thus,
the solution,   $\Psi$,  is given by
\bea
\label{psi}
\Psi (a,\alpha ,\beta)={\cal U}_{\nu}(a)e^{im\beta}P^m_j({\rm cosh}\alpha)
\eea
In quantum cosmology one chooses suitable linear combinations of the
solutions to derive the wave function of the Universe,   depending
on the boundary conditions such as the one adopted by Hartle and Hawking
\cite{hh} (the no boundary proposal) or one proposed by Vilenkin \cite{av}.
 In the context of
string cosmology, interesting classical solutions are derived in the pre-big
bang scenario \cite{gv}; alterntively in the big crunch to big bang approach
\cite{bcbb}.
These initial conditions have interesting consequences in the studies
of quantum string cosmologies \cite{cop1,mmp}.
In this note, we shall not persue
to derive  the wave function of the Universe, in our approach,   according to
various proposals in quantum cosmology and differ this aspect of
quantum string cosmology to a separate
investigation. \\ 
We proceed to unravel a  novel symmetry which, 
in our openion, is 
quite interesting and is unexplored in axion-dilaton  string cosmology. 
The {\it raison de etre} of the symmetry is the invariance of the theory
under the S-duality group, $SU(1,1)$ and the wave functions (\ref{psi})
are endowed with  huge
degeneracy, since $m$ is unbounded (from below or above depending on the 
choice of wave function). This is an attribute of the noncompact nature
of $SU(1,1)$.\\   
Notice that  one may  construct tensor operators,
from the set $\{J_i \}$,  which
transform as higher dimensional representation of $SU(1,1)$. Such 
operators act in a Hilbert space, where the Casimir operator takes a specific
value, say $C=\lambda$, and the operators close into an infinite dimensiosal
algebra. This algebra is parametrized  by $\lambda$ and is denoted  as 
${\cal T} (\lambda)$. Such an approach has been persued by Pope etal.\cite{cp}
 for
$SL(2,R)$ group, in detail,  to study $W_{\infty}$ algebra 
in an abstract setting. 
In the present context, the wave functions are known explicitly and the
operators may be constructed in terms of $J_3 , J_{\pm}$ as given below 

\bea
\label{jop}
J_{\pm}={\mp}e^{\pm i\beta}{{\partial }\over {\partial \alpha}} -i{\rm coth}
\alpha e^{\pm i\beta}{{\partial }\over {\partial \beta}},~J_3=-i{{\partial}
\over {\partial \beta}}
\eea
Then one might adopt the prescription of 
ref.\cite{cp}.\\
As an illustative example,
let us consider a specific case,  
$C=-3/16$, to show how the $w_{\infty}$ algebra emerges. A more elegant
way is to express the generators  
in terms of a single boson creation and 
annihilation operators \cite{ui} which serve our purpose better.
\bea
\label{oneb}
J_{+}={1\over 2}(a^{\dagger})^2 ,
 ~J_-={1\over 2}a^2 ,
~ J_3={1\over 2}(a^{\dagger}a
+1/2)
\eea
$[a, a^{\dagger }]=1$. Defining,
$|n>=(n!)^{-1/2}(a^\dagger)^n|0>$, where $a|0>=0$ is the condition on vacuum,
we find
\bea
\label{raise}
J_+|n>={{\sqrt {(n+1)(n+2)}}\over 2}|n+2>, 
\eea
\bea
J_-|n>={{\sqrt {(n(n-1)}}\over 2}|n-2>, 
\eea
and
\bea
 J_3|n>=(n+{1\over 2})|n>
\eea
We get two
different representations of the Lie algebra 
from  $|n>$: (i) For odd $n$,  $j=-3/4$ 
and (ii) for even $n$, $j=-1/4$ which belongs to the supplementary series.
When $J_3$ and $J_{\pm}$ are defined 
as in (\ref{oneb}),
a  suitable set of operators can be constructed, in both the cases, 
which satisfy the $w_{\infty}$ algebra \cite{bakas}
\bea
\label{w}
[{\cal T}_{n,m} ,{\cal T}_{k,l}] &=& \left((m+1)(k+1)-(n+1)(l+1)\right)
{\cal T}
_{n+k,m+l}
\eea
where ${\cal T}_{n,m}=(a^{\dagger})^{n+1}a^{m+1},~n,m \ge -1$. Here,
 we have computed  the classical algebra, ignoring  
normal orderings.\\   
In what follows, we briefly discuss the procedure
 to construct \cite{cp} the tensor operators satisfying
the algebra ${\cal T}(\lambda)$ for arbitrary $\lambda$. 
First,
one starts with a highest weight tensor, say $T^l=(J_+)^l$ and then 
obtains lower
weight operators by successive actions of $J_-$. Such objects commute
with the Casimir. As remarked earlier, for a given $\lambda$, ${\cal T}^l_m$
close into an infinite dimensional algebra.
An explicit construction is   
\bea
\label{tensors}
{\cal T}^l_m ={\cal N}(Ad_{{J{_-}}})^{l-m}(J_+)^l
\eea 
with the definition $Ad_X(Y)=[X,Y]$ and $\cal N$ is a suitable normalization
constant. These tensors can be expressed as a product of polynomial
of order $l-m$ in  $J_3$ and $J_+ ^l$, since $[J_- ,(J_+)^l]=2lJ_3(J_+)^{l-1}
-l(l-1)(J_+)^{l-1} $. The representation (\ref{jop}) may be used to realize  
the operators defined in  (\ref{tensors}). Thus, for a given $\lambda$, one 
gets the $W_{\infty}(\lambda)$ algebra. Note that for 
special value of $\lambda =0$,
one obtains the subalgebra $W_{\wedge}$ which is contained in $W_{\infty}$,
  as has been elaborated in \cite{cp}.
The former still allows an infinite dimensional representation.   
In our cosmological scenario, the operators $J_{\pm}$ act on the wave function
obtained from solutions of differtial equation associate with $SU(1,1)$ Casimir
(\ref{lb}). Thus, these operators acting on $\Psi$ produce another wave
 function, with same $j(j+1)$ (alternatively same $\lambda$), but still
it is a solution of WDW equation. In other words, the energy eigenvalue is still
zero for such a wave function. \\   
To summarize,  we have derived hitherto unknown solutions to WDW equation in
axion-dilaton cosmology which are highly degenerate. The implications of the  
such degeneracies in string quantum cosmology 
are not completely explored. 
The existence
of $W$-infinity symmetry is a surprize in
the present  investigation of string cosmology   
which originates from the underlying S-duality
invariance of the theory. String cosmology with dilaton alone is unlikely to  
to possess such a symmetry.  
It is  very tempting to ask whether the symmetries presented here have any
role to address the cosmological constant problem, since it is hoped
that string theory, with its rich symmetry contents, might provide 
a resolution of the cosmological  constant problem. \\ 
I have benefited from discussions with A. Dhar, Z. F. Ezawa and C. Zachos
at various stages of this work. I would like to thank Werner R\"uhl for useful 
correspondence. The gracious  hospitality of Yoshihita Kitazawa and KEK is
acknowledged.

%%%%%%%%%%%%%%%%%%%%%%%%%%%%%%%%%%%%%%%%%%%%%%%%%%%%%%%%%%%%%%%%%%%%%%

%ecenterline{{\bf References}}
%\begin{enumerate}

%%%%%%%%%%%%%%%%%%%%%%%%%%%%%%%%%%%%%%%%%%%%%%%%%%%%%%%%%%%%%%
\end{document}